# Stabilized microwave-frequency transfer using optical phase sensing and actuation

SASCHA W. SCHEDIWY,[1,*] DAVID R. GOZZARD,[1] SIMON STOBIE[1], J.A. MALAN[2], AND KEITH GRAINGE[3]

[1]*International Centre for Radio Astronomy Research, School of Physics and Astrophysics, University of Western Australia, Perth, WA 6009, Australia*
[2]*SKA SA, Park Road, Pinelands, 7405, South Africa*
[3]*Jodrell Bank Observatory, University of Manchester, M13 9PL, United Kingdom*
*Corresponding author: sascha.schediwy@uwa.edu.au*

**We present a stabilized microwave-frequency transfer technique that is based on optical phase-sensing and optical phase-actuation. This technique shares several attributes with optical-frequency transfer and therefore exhibits several advantages over other microwave-frequency transfer techniques. We demonstrated stabilized transfer of an 8,000 MHz microwave-frequency signal over a 166 km metropolitan optical fiber network, achieving a fractional frequency stability of $6.8\times10^{-14}$ Hz/Hz at 1 s integration, and $5.0\times10^{-16}$ Hz/Hz at $1.6\times10^{4}$ s. This technique is being considered for use on the Square Kilometre Array SKA1-mid radio telescope.**

http://dx.doi.org/10.1364/OL.99.099999

Stabilized frequency transfer over optical fiber is rapidly moving from experimental demonstrations, to actively supporting a broad range of real-world applications. The frequency of the transmitted signal – optical, radio, or microwave – dictates which applications can be supported with this technology.

Stabilized optical-frequency transfer techniques lead this research in terms of demonstrated transfer distance [1], frequency stability performance [2], and network roll-out [3]. However, most practical applications, including radio astronomy, geodesy, finance, navigation, defense, and aspects of fundamental physics [4,5], require transfer of stabilized radio- or microwave frequency signals to directly interface with the application's electronic systems. Optical combs can be used to translate optical frequencies into the radio or microwave domain [6], but they remain too complex, bulky, or costly for many applications. Therefore, most practical support of the aforementioned applications involves microwave-frequency transfer techniques, but these are constrained in the following ways:

The standard stabilized microwave-frequency transfer techniques [7] require group-delay actuation to compensate the physical length changes of the fiber link. For practical deployments over long links, this usually involves implementing a combination of fiber stretcher (medium actuation speed and very limited range) in series with a thermal spool (slow actuation speed and physically bulky). In contrast, the acousto-optic modulator (AOM) actuators used in stabilized optical-frequency transfer are capable of the faster actuation speeds, as well as having infinite feedback range. Radio-frequency phase conjugation techniques [8] have been demonstrated over longer distances [9,10] than standard stabilized microwave-frequency transfer, but have yet to match their transfer performance.

In addition, stabilized radio- or microwave-frequency transfer techniques require the returned signal to be rebroadcast at either a different modulation frequency, optical wavelength, or fiber core, to avoid frequency overlap from unwanted reflections on the link, which can cause the servo to function improperly. These reflection mitigating methods then can bring about additional complications, including those resulting from optical polarization and chromatic dispersion, which in turn requires further complexity. In stabilized optical-frequency transfer, strategically placed AOMs can be used to simply apply static optical-frequency shifts [11] to avoid these issues.

In this Letter we report on a stabilized microwave-frequency transfer technique that is based on optical phase sensing and actuation, and which is therefore able to utilize many of the key advantages of stabilized optical transfer.

As shown in Figure 1, an optical signal with frequency $\nu_L$, generated by a laser at the **Local Site**, is injected into two arms of a Mach-Zehnder interferometer (MZI). A dual-parallel Mach-Zehnder modulator (DPM) is located in Arm 1 of the MZI, and a microwave frequency of $\nu_{DP}$ is applied to the DPM electronic inputs. The phase of the electronic inputs, and the DPM voltage biases, are tuned to generate single-sideband suppressed-carrier (SSB-SC) modulation, thereby producing a static microwave-frequency shift $\nu_{DP}$ on the optical signal. Arm 2 of the MZI contains an acousto-optic modulator (AOM), which adds the servo AOM radio-frequency shift $\nu_{A\text{-}srv}$ and the servo actuation signal $\Delta\nu_{A\text{-}srv}$ to the optical signal. In addition, the optical signals in each of the two arms of the MZI pick up undesirable non-common phase noise $\Delta\phi_{MZI,i}$ (where $i$ is the index representing the optical signals in Arm 1 and Arm 2 of the MZI).

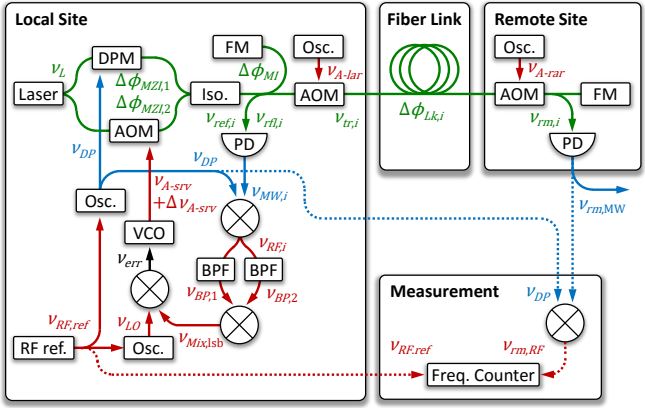

**Fig. 1.** Schematic diagram of our stabilized microwave-frequency transfer technique. Optical-frequency signals are shown in green; microwave-frequency signals in blue; radio-frequency signals in red; and the error signal in black. DPM dual-parallel Mach-Zehnder modulator; AOM acousto-optic modulator; Osc. Oscillator; Iso. optical isolator; FM Faraday mirror; PD photodetector; VCO voltage-controlled oscillator; BPF band-pass filter; and RF ref. radio-frequency reference.

Just as is the case in standard stabilized optical transfer techniques [2], the optical signals then enter a Michelson interferometer (MI) via an optical isolator (to prevent reflections returning to the laser). The **Fiber Link** is incorporated into the long arm of the MI, with the short arm providing the physical reference for the optical phase sensing. The optical reference signals $\nu_{ref,i}$ at the photodetector are

$$\nu_{ref,1} = \nu_L + \nu_{DP} + \tfrac{1}{2\pi}\big(\Delta\dot\phi_{MZI,1} + 2\Delta\dot\phi_{MI}\big), \text{and} \quad (1)$$

$$\nu_{ref,2} = \nu_L + (1+\Delta)\nu_{A\text{-}srv} + \tfrac{1}{2\pi}\big(\Delta\dot\phi_{MZI,2} + 2\Delta\dot\phi_{MI}\big), \quad (2)$$

where $\Delta\phi_{MI}$ is the undesirable phase noise picked up by the optical signals passing through the MI reference arm.

A 'local anti-reflection' AOM that applies a static frequency offset of $\nu_{A\text{-}lar}$ can be incorporated into either the long arm (as shown in Figure 1) or the reference arm of the MI. This allows the servo electronics to distinguish $\nu_{ref,i}$ from unwanted reflections on the link. (Note; the anti-reflection AOMs are not critical for the technique, but are useful for practical implementation on fiber links that may contain unwanted reflections.) With the local anti-reflection AOM placed in the long arm, the optical signals transmitted across the **Fiber Link** $\nu_{tr,i}$ are

$$\nu_{tr,1} = \nu_L + \nu_{DP} + \nu_{A\text{-}lar} + \tfrac{1}{2\pi}\Delta\dot\phi_{MZI,1}, \text{and} \quad (3)$$

$$\nu_{tr,2} = \nu_L + (1+\Delta)\nu_{A\text{-}srv} + \nu_{A\text{-}lar} + \tfrac{1}{2\pi}\Delta\dot\phi_{MZI,2}. \quad (4)$$

As these signals are transmitted across the fiber link, they pick-up phase noise $\Delta\phi_{Lk,i}$ from random optical path length changes in the link that are unique to their specific transmitted frequency.

At the **Remote Site**, the two optical signals pass through a remote anti-reflection AOM with a static frequency of $\nu_{A\text{-}rar}$ to produce the following two remote signals $\nu_{rm,i}$:

$$\nu_{rm,1} = \nu_L + \nu_{DP} + \nu_{A\text{-}lar} + \nu_{A\text{-}rar} + \tfrac{1}{2\pi}\big(\Delta\dot\phi_{MZI,1} + \Delta\dot\phi_{Lk,1}\big), \quad (5)$$

$$\nu_{rm,2} = \nu_L + (1+\Delta)\nu_{A\text{-}srv} + \nu_{A\text{-}lar} + \nu_{A\text{-}rar} + \tfrac{1}{2\pi}\big(\Delta\dot\phi_{MZI,2} + \Delta\dot\phi_{Lk,2}\big). \quad (6)$$

At the remote site, the signal is split into two fiber paths, with one set of signals going to a photodetector and the other to a Faraday mirror. At the photodetector the beat between the two optical signals $\nu_{rm,1}$ and $\nu_{rm,2}$ is recovered as the **Remote Site** microwave-frequency electronic signal

$$\nu_{rm,MW} = (1+\Delta)\nu_{A\text{-}srv} - \nu_{DP} + \tfrac{1}{2\pi}\big(\Delta\dot\phi_{MZI,2} - \Delta\dot\phi_{MZI,1} + \Delta\dot\phi_{Lk,2} - \Delta\dot\phi_{Lk,1}\big). \quad (7)$$

On the other fiber path, a Faraday mirror reflects the two optical signals back to the **Local Site** across the **Fiber Link**, with each signal receiving additional optical shifts $\nu_{A\text{-}rar}$ and $\nu_{A\text{-}lar}$ when passing through the remote and local AOMs a second time. In addition, the two optical signals pick up another copy of $\Delta\phi_{Lk,i}$ from the **Fiber Link**. The two reflected optical signals $\nu_{rfl,i}$ then strike the servo photodetector to produce

$$\nu_{rfl,1} = \nu_L + \nu_{DP} + 2(\nu_{A\text{-}lar} + \nu_{A\text{-}rar}) + \tfrac{1}{2\pi}\big(\Delta\dot\phi_{MZI,1} + 2\Delta\dot\phi_{Lk,1}\big), \text{and} \quad (8)$$

$$\nu_{rfl,2} = \nu_L + (1+\Delta)\nu_{A\text{-}srv} + 2(\nu_{A\text{-}lar} + \nu_{A\text{-}rar}) + \tfrac{1}{2\pi}\big(\Delta\dot\phi_{MZI,2} + 2\Delta\dot\phi_{Lk,2}\big). \quad (9)$$

The mixing of the two reference frequencies $\nu_{ref,i}$ and the two reflected optical frequencies $\nu_{rfl,i}$ results in six primary electronic mixing products $\nu_{MW,j}$ (where $j$ is the index 1 to 6). Of those, the two microwave-frequency signal mixing products that are crucial for this technique are

$$\nu_{MW,1} = \nu_{ref,1} - \nu_{rfl,2} = \nu_{DP} - (1+\Delta)\nu_{A\text{-}srv} - 2(\nu_{A\text{-}lar} + \nu_{A\text{-}rar}) + \tfrac{1}{2\pi}\big(\Delta\dot\phi_{MZI,1} - \Delta\dot\phi_{MZI,2}\big) + \tfrac{1}{\pi}\big(\Delta\dot\phi_{MI} - \Delta\dot\phi_{Lk,2}\big), \text{and} \quad (10)$$

$$\nu_{MW,2} = \nu_{ref,2} - \nu_{rfl,1} = (1+\Delta)\nu_{A\text{-}srv} - \nu_{DP} - 2(\nu_{A\text{-}lar} + \nu_{A\text{-}rar}) + \tfrac{1}{2\pi}\big(\Delta\dot\phi_{MZI,2} - \Delta\dot\phi_{MZI,1}\big) + \tfrac{1}{\pi}\big(\Delta\dot\phi_{MI} - \Delta\dot\phi_{Lk,1}\big). \quad (11)$$

Given an appropriate selection of AOM frequencies, the other four mixing products (and all intermodulations) occur at different frequencies.

Once in the electronic domain, the microwave-frequency signals $\nu_{MW,j}$ can be mixed with a copy of $\nu_{DP}$, to produce the following two critical electronic radio-frequency signals:

$$\nu_{RF,1} = -(1+\Delta)\nu_{A\text{-}srv} - 2(\nu_{A\text{-}lar} + \nu_{A\text{-}rar}) + \tfrac{1}{2\pi}\big(\Delta\dot\phi_{MZI,1} - \Delta\dot\phi_{MZI,2}\big) + \tfrac{1}{\pi}\big(\Delta\dot\phi_{MI} - \Delta\dot\phi_{Lk,2}\big), \text{and} \quad (12)$$

$$\nu_{RF,2} = (1+\Delta)\nu_{A\text{-}srv} - 2(\nu_{A\text{-}lar} + \nu_{A\text{-}rar}) + \tfrac{1}{2\pi}\big(\Delta\dot\phi_{MZI,2} - \Delta\dot\phi_{MZI,1}\big) + \tfrac{1}{\pi}\big(\Delta\dot\phi_{MI} - \Delta\dot\phi_{Lk,1}\big). \quad (13)$$

The electronic signal path is then split, with each path containing a band-pass filter that is centered on one of the above frequencies. These filters reject the opposing signal, as well as the other unwanted mixing products ($\nu_{MW,3}$ to $\nu_{MW,6}$) and any frequency intermodulations. The signals $\nu_{RF,1}$ and $\nu_{RF,2}$ are then mixed to produce the lower-sideband

$$\nu_{Mix,lsb} = \nu_{RF,2} - \nu_{RF,1} = 2\Big((1+\Delta)\nu_{A\text{-}srv} + \tfrac{1}{2\pi}\big(\Delta\dot\phi_{MZI,2} - \Delta\dot\phi_{MZI,1} + \Delta\dot\phi_{Lk,2} - \Delta\dot\phi_{Lk,1}\big)\Big). \quad (14)$$

A low-pass filter is used to reject the upper-sideband and other products. Finally, $\nu_{Mix,\mathrm{lsb}}$ is mixed with the servo local oscillator $\nu_{LO}$ (set to $2\nu_{A\text{-}srv}$) to produce the servo error signal

$$\nu_{err} = 2\left(\Delta\nu_{A\text{-}srv} + \tfrac{1}{2\pi}(\Delta\dot{\phi}_{MZI,2} - \Delta\dot{\phi}_{MZI,1} + \Delta\dot{\phi}_{Lk,2} - \Delta\dot{\phi}_{Lk,1})\right). \quad (15)$$

The servo error signal is applied to a voltage controlled oscillator (VCO) with a nominal frequency $\nu_{A\text{-}srv}$. The VCO output goes to the servo AOM, thereby closing the servo loop. When the servo is engaged, $\nu_{err}$ is driven to zero so

$$\Delta\nu_{A\text{-}srv} = -\tfrac{1}{2\pi}(\Delta\dot{\phi}_{MZI,2} - \Delta\dot{\phi}_{MZI,1} + \Delta\dot{\phi}_{Lk,2} - \Delta\dot{\phi}_{Lk,1}). \quad (16)$$

Substituting this into Equation 7 shows that the undesirable non-common phase noise picked up in the **Fiber Link**, as well as the phase noise from the MZI in the **Local Site**, is canceled out (within the light round-trip bandwidth and other practical gain limitations). This gives

$$\nu^*_{rm,\mathrm{MW}} = \nu_{A\text{-}srv} - \nu_{DP}, \quad (17)$$

where $\nu^*_{rm,\mathrm{MW}}$ is the MW remote signal with the servo engaged.

We describe the experimental verification of this method using 8,000 MHz transfer, with all optical elements fiberized. An *NKT Photonics* Koheras BASIK X15 laser (spectral linewidth <100 Hz) was used to produce an optical frequency of $\nu_L = 193$ THz (corresponding to a wavelength of 1552 nm). The DPM was a *Photline* MXIQ-LN-40 configured to produce a down-shift of the optical-frequency by $\nu_{DP} = -7{,}960$ MHz. The microwave-frequency oscillator was an *Agilent* N5183A MXG, with all other radio-frequency signals supplied by a *Liquid Instruments* Moku:Lab.

The AOMs were *IntraAction* FCM-series, with $\nu_{A\text{-}srv} = +40$ MHz, resulting in a frequency difference between $\nu_{tr,1}$ and $\nu_{tr,2}$ of 8,000 MHz. The anti-reflection AOMs had frequencies of $\nu_{A\text{-}lar} = +50$ MHz, and $\nu_{A\text{-}rar} = +50$ MHz. For the experiment described here, the local anti-reflection AOM was located in the link arm of the MI as shown in Figure 1.

Polarization maintaining fiber was used until the MZI output to ensure the polarization into the DPM was optimally aligned, and that the MZI optical power remained maximized. The signal was transmitted through installed metropolitan optical fiber networks up to 166 km in length (2× 83 km *AARNet*-managed fiber loops). This resulted in a two-way light round trip time of 1.6 ms, limiting the servo bandwidth to around 600 Hz. The total optical loss of the fiber link was 47 dB. Two *IDIL Fibres Optiques* bi-directional optical amplifiers were used to boost signal strength, with one amplifier located after the first 83 km fiber loop, and the other just prior to the **Local Site**. *Discovery Semiconductors* DSC-R402 and DSC-R401HG photodetectors were used for the opto-electronic conversion. Faraday mirrors were used at the ends of the two arms of the MI to ensure the signals reflected back to the servo photodetector at the **Local Site** were aligned in polarization.

The combination of the AOM frequencies used in this experiment resulted in the following two key microwave-frequency electronic signals at the output of the servo photodetector; $\nu_{MW,1} = 8{,}200$ MHz and $\nu_{MW,2} = 7{,}800$ MHz. After mixing with $\nu_{DP}$ this resulted in the following radio-frequency signals; $\nu_{RF,1} = 240$ MHz and $\nu_{RF,2} = 160$ MHz. The mix of these two signals produced the lower-sideband $\nu_{Mix,\mathrm{lsb}} = 80$ MHz. The local oscillator $\nu_{LO}$ was set to 80 MHz to produce a DC error signal $\nu_{err}$ upon mixing with $\nu_{Mix,\mathrm{lsb}}$.

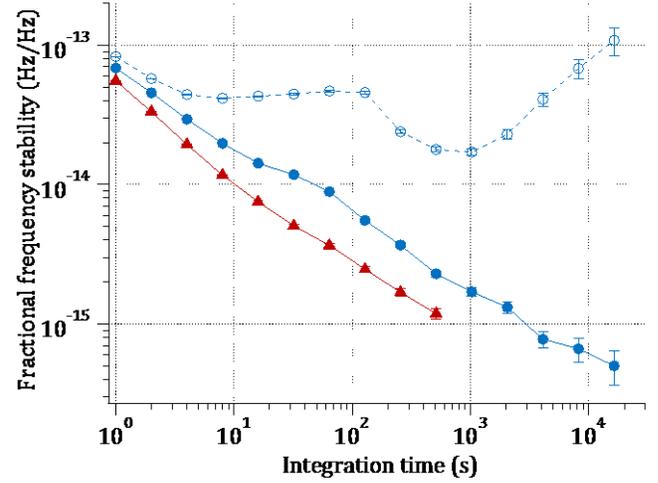

**Fig. 2**. Fractional frequency stability of 8,000 MHz transfer over a 166 km metropolitan optical fiber network (blue color, circle markers) with the stabilization servo engaged (solid line, filed markers) and servos disengaged (dashed line, open markers). A stabilized transfer over a 2 m fiber patch lead is shown by the red triangle markers.

Both the **Local Site** and **Remote Site** were co-located in the same laboratory enabling an independent out-of-loop measurement of the transfer stability, with all electronic equipment referenced to an *IEM-KVARZ* CH1-75A active hydrogen maser. The 8,000 MHz **Remote Site** microwave-frequency electronic signal was mixed with $\nu_{DP}$ to produce a radio-frequency signal at $\nu_{rm,RF} = 40$ MHz that could be directly probed by an *Agilent* 53132A high-precision frequency counter (gate time set to 1 s). The output data was used to produce a triangle weighted estimate of the frequency stability.

Figure 2 shows the fractional frequency stability of three 8,000 MHz transfer measurements plotted as a function of integration time. The blue dashed line with open markers shows the unstabilized transfer over 166 km. Here the transfer stability is $8.4\times10^{-14}$ Hz/Hz at 1 s of integration, and $1.1\times10^{-13}$ Hz/Hz at $1.6\times10^4$ s. The blue solid line with solid markers represents transfer with the stabilization servo engaged. At 1 s of integration the stability is $6.8\times10^{-14}$ Hz/Hz, and at $1.6\times10^4$ s it is $5.0\times10^{-16}$ Hz/Hz; demonstrating a suppression of fiber noise by more than two orders-of-magnitude. The noise floor of the transfer system, as measured by stabilized transfer over a 2 m fiber patch lead, is displayed as the red solid line with triangle markers. It starts at a value of $5.5\times10^{-14}$ Hz/Hz at 1 s, and drops to $1.2\times10^{-15}$ Hz/Hz by 512 s. In addition, the 166 km stabilized frequency transfer was shown to be accurate to within 20 µHz.

We have described and experimentally demonstrated the efficacy of a stabilized microwave-frequency transfer technique that is based on optical phase sensing and optical phase actuation. While it does not achieve the same level of ultimate transfer stability performance as the world-leading results [7,12], it does exhibits several advantages over other radio- or microwave-frequency transfer techniques, as discussed in the following six paragraphs below.

The use of AOMs enables the construction of compact servo systems at the **Local Site** by not requiring bulky fiber stretchers or thermal spools. The use of the AOMs eliminates feedback range concerns due to their infinite feedback range; and, therefore, the servo systems can be constructed without potentially complex integrator-reset circuits. Finally, AOMs have faster feedback than fiber stretchers or thermal spools, which can allows for optimized servo gains for short links that are not limited by the light round-trip time.

The technique actively suppresses phase noise originating not just from the **Fiber Link** in the MI arm, but also from the MZI at the **Local Site**. Plus, given the simplicity of components at the **Remote Site**, the technique is largely immune to environmental perturbations on the system's hardware.

As in stabilized optical transfer, anti-reflection AOMs can be used to generate appropriate optical-frequency shifts to mitigate unwanted reflections that are present on most real-world links. Our technique therefore requires only a single laser, reducing system complexity. Further system optimization can be achieved by locating the local anti-reflection AOM in the MI reference arm, which has the benefit of removing some unnecessary optical loss from the link arm.

The technique utilizes Faraday mirrors at the ends of the MI arms to give maximum detected signal at the servo photodetector, as is done with stabilized optical transfer. This removes the need for any initial polarisation alignment, or any ongoing polarization control or polarization scrambling. The technique can be deployed on standard fiber links alongside data transmission and does not require specialty fiber, such as dispersion compensation or polarization maintaining fiber, in the **Fiber Link**.

The microwave signal being transmitted on the **Fiber Link** arises from only two optical signals, not three as is the case for standard intensity modulation commonly used in radio- or microwave-frequency transfer. Using only two optical signals ensures that that the maximum signal power is available at the **Remote Site** regardless of link length. This also enables the transmission frequency to be varied without consideration of link length.

Bi-directional optical amplifiers can be deployed to extend the range of transmission, and, therefore, potentially complex electronic signal re-generation systems are not required. Two optical amplifiers were used to demonstrate stabilized microwave-frequency transfer over a 166 km metropolitan optical fiber network, around twice the distance of previously published microwave-frequency (>1 GHz) transfer [7,12].

This stabilized microwave-frequency transfer technique is one of two being considered for selection as the phase-synchronization system for the Square Kilometre Array SKA1-mid radio telescope [13,14]. In addition, we are exploring the technique's role in the MeerKAT radio telescope [15] as part of a potential future X-band receiver upgrade.

**Funding.** University of Manchester; University of Western Australia (UWA).

**Acknowledgment.** The authors wish to thank *AARNet* for the provision of light-level access to their fiber network infrastructure. This paper describes work being carried out for the SKA Signal and Data Transport (SaDT) consortium as part of the Square Kilometre Array (SKA) project. The SKA project is an international effort to build the world's largest radio telescope, led by SKA Organisation with the support of 10 member countries.